\documentclass[english,]{pandoc/chi-ext}
\usepackage[T1]{fontenc}
\usepackage{lmodern}
\usepackage{amssymb,amsmath}
\usepackage{ifxetex,ifluatex}
\usepackage{fixltx2e} 
\IfFileExists{upquote.sty}{\usepackage{upquote}}{}
\ifnum 0\ifxetex 1\fi\ifluatex 1\fi=0 
  \usepackage[utf8]{inputenc}
\else 
  \usepackage{fontspec}
  \ifxetex
    \usepackage{xltxtra,xunicode}
  \fi
  \defaultfontfeatures{Mapping=tex-text,Scale=MatchLowercase}
  
\fi
\IfFileExists{microtype.sty}{\usepackage{microtype}}{}
\usepackage[numbers]{natbib}
\usepackage{graphicx}
\makeatletter
\def\maxwidth{\ifdim\Gin@nat@width>\linewidth\linewidth
\else\Gin@nat@width\fi}
\makeatother
\let\Oldincludegraphics\includegraphics
\renewcommand{\includegraphics}[1]{\Oldincludegraphics[width=\maxwidth]{#1}}
\ifxetex
  \usepackage{hyperref}
\else
  \usepackage{hyperref}
\fi
\hypersetup{breaklinks=true,
            bookmarks=true,
            pdfauthor={},
            pdftitle={},
            colorlinks=true,
            urlcolor=blue,
            linkcolor=magenta,
            pdfborder={0 0 0}}
\ifxetex
  \usepackage{polyglossia}
  \setmainlanguage{english}
\else
  \usepackage[english]{babel}
\fi

\usepackage{balance}
\usepackage{pandoc/bibspacing}

\usepackage{comment}

\usepackage{marginnote}

\author{}
\date{}

\begin{document}

\excludecomment{quote}

\let\oldfootnote\footnote

\renewcommand{\footnote}[1]
{
  \marginpar{#1}
}

\def\plaintitle{Assessing the Zone of Comfort in Stereoscopic Displays using EEG}
\def\plainauthor{Jérémy Frey}
\def\plainkeywords{Stereoscopy; Comfort; EEG; HCI evaluation}

\title{\plaintitle}

\copyrightinfo{
  Copyright 2014 held by Owner/Author. Publication Rights Licensed to ACM.
}

\copyrightinfo{\scriptsize Permission to make digital or hard copies of part or all of this work for personal or classroom use is granted without fee provided that copies are not made or distributed for profit or commercial advantage and that copies bear this notice and the full citation on the first page. Copyrights for third-party components of this work must be honored. For all other uses, contact the owner/author(s). Copyright is held by the author/owner(s). \\
{\emph{CHI 2014}}, April 26--May 1, 2014, Toronto, Ontario, Canada. \\
ACM 978-1-4503-2474-8/14/04. \\
http://dx.doi.org/10.1145/2559206.2581191}

\numberofauthors{4} \author{
  \alignauthor{
    \textbf{Jérémy Frey}$^{1,2,3}$\\
    \email{jeremy.frey@inria.fr}
  }\alignauthor{
    \textbf{Léonard Pommereau}$^{3}$\\
    \email{leonard.pommereau@inria.fr}
  }
  \vfill
  \alignauthor{
    \textbf{Fabien Lotte}$^{3,1,2}$\\
    \email{fabien.lotte@inria.fr}
  }\alignauthor{
    \textbf{Martin Hachet}$^{3,1,2}$\\
    \email{martin.hachet@inria.fr}
  }
  \vfill
}

\teaser{

$^{1}$Univ. Bordeaux, LaBRI, UMR 5800, F-33400 Talence, France. \\
$^{2}$CNRS, LaBRI, UMR 5800, F-33400 Talence, France. \\
$^{3}$INRIA, F-33400 Talence, France. \\
\vfill

}

\hypersetup{
  pdftitle={\plaintitle},
  pdfauthor={\plainauthor},  
  pdfkeywords={\plainkeywords},
  citecolor=black,
  linkcolor=blue,
  menucolor=black,
  urlcolor=blue,
}

\maketitle

\begin{abstract}

The conflict between vergence (eye movement) and accommodation (crystalline lens deformation) occurs in every stereoscopic display. It could cause important stress outside the "zone of comfort",  when stereoscopic effect is too strong. This conflict has already been studied using questionnaires, during viewing sessions of several minutes. The present pilot study describes an experimental protocol which compares two different comfort conditions using electroencephalography (EEG) over short viewing sequences. Analyses showed significant differences both in event-related potentials (ERP) and in frequency bands power. An uncomfortable stereoscopy correlates with a weaker negative component and a delayed positive component in ERP. It also induces a power decrease in the alpha band and increases in theta and beta bands. With fast responses to stimuli, EEG is likely to enable the conception of adaptive systems, which could tune the stereoscopic experience according to each viewer.

\end{abstract}



\section{Introduction}\label{introduction}

Stereoscopy is a technique which gives to viewers the illusion of depth
by sending a different image to each eye. Stereoscopy has been studied
for decades as a way to improve the perception of 3D scenes. Even so, it
is only circa 2010 that the use of stereoscopic displays has been
popularized among the general public, \emph{e.g.}, in movies or video
gaming industries. We investigate a new, innovative, methodology to
assess comfort in stereoscopy.\footnote{\begin{figure}[htbp]
  \centering
  \includegraphics{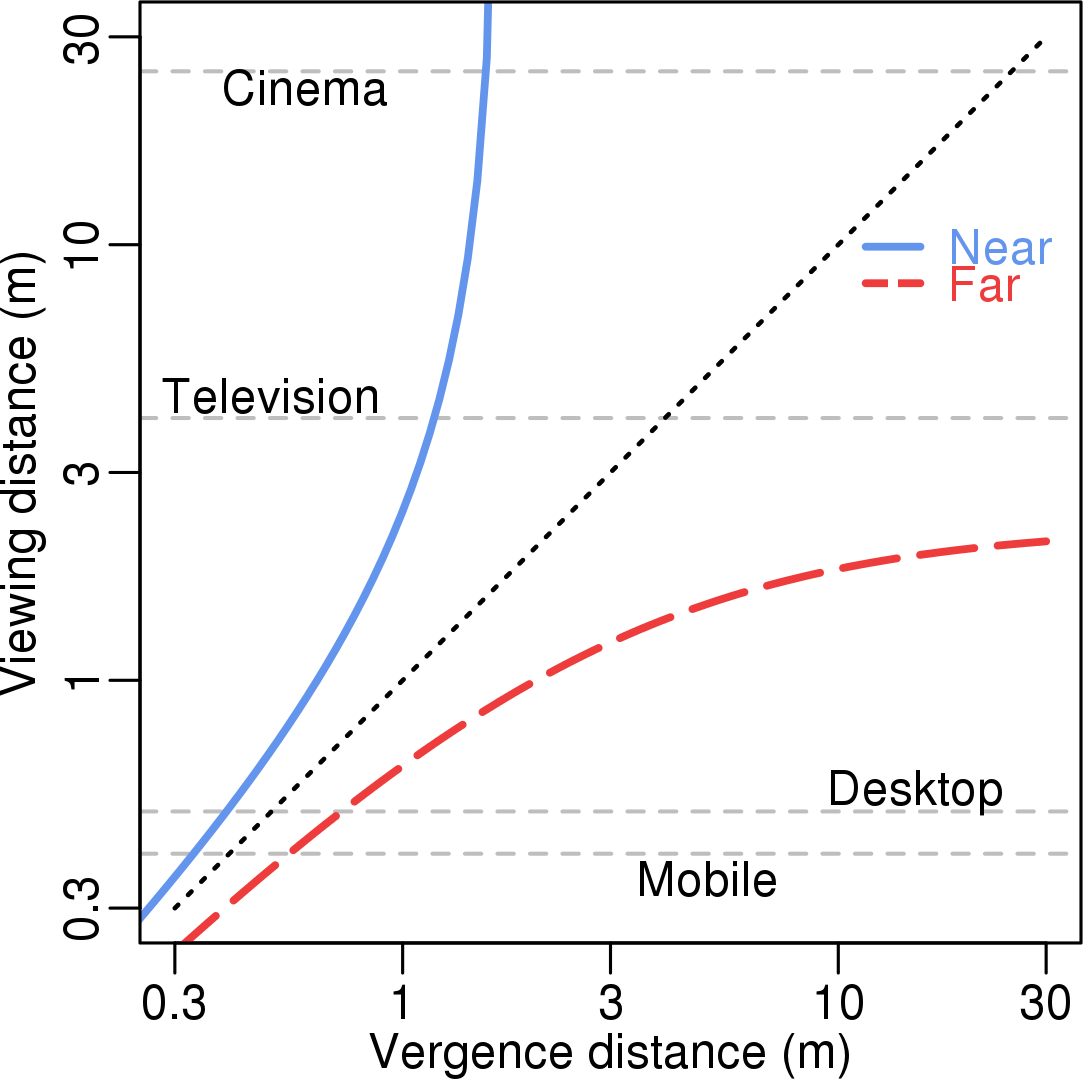}
  \caption{The acceptable zone of comfort depending on viewing distance
  and vergence distance (\emph{i.e.} the apparent depth of contents).
  From \citep{Shibata2011}.\label{comfort_zone_fig}}
  \end{figure}}

\begin{figure}[htbp]
\centering
\includegraphics{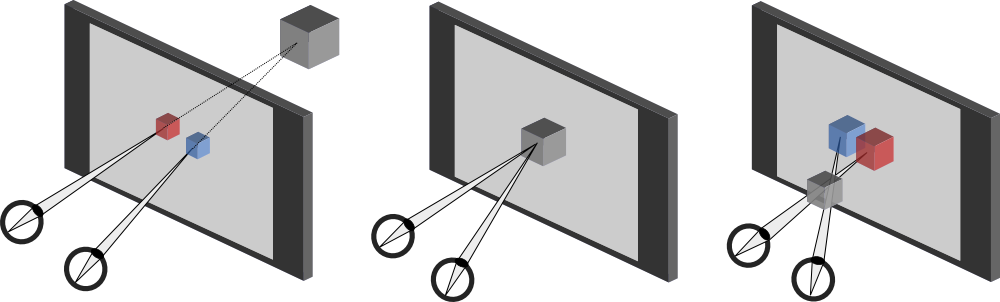}
\caption{The vergence-accommodation conflict (VAC). \textbf{Left}:
object ``behind'' the screen, negative VAC. \textbf{Middle}: the object
appears flat, no VAC. \textbf{Right}: object ``in front'', positive
VAC.\label{VAC_fig}}
\end{figure}

This study focuses on the first and foremost parameter tied to
stereoscopy: apparent depth. In stereoscopic vision, the depth
perception of an object is due to the discrepancy between its projection
on both retinas. In order to see \emph{one} object, the eyeballs need to
rotate accordingly. This mechanism is called the ``vergence''. In a
natural situation, as an object is moving closer or further, vergence
matches another physiological phenomena: ``accommodation''. It enables
the object's image to remain clear on the retina. It is caused by a
deformation of the crystalline lens, which focuses light beams the same
way camera lenses do. With stereoscopic displays however, the physical
distance of the screen -- thus the accommodation -- remains still while
apparent depth and vergence vary. This unecological situation causes a
stress on viewers. It is called the ``vergence-accommodation conflict''
(VAC, see figure \ref{VAC_fig}). The VAC is inextricably linked to every
type of stereoscopic displays and is one of the major causes of
discomfort \citep{Shibata2011}; this is why we decided to manipulate
this variable in our investigations.

On the one hand the VAC has already been studied using questionnaires;
in \citep{Shibata2011} a ``zone of comfort'' has been established
(figure \ref{comfort_zone_fig}). The apparent depth of objects should
remain within this zone for the viewers not to feel discomfort. On the
other hand electroencephalography (EEG) has been used to study fatigue
induced by stereoscopic displays \citep{Li2008a, Cho2012, Chen2013}.
While results related to the VAC are great insights on how to present
content to viewers, these are general rules based on controlled
experimental setups. The match cannot be perfect with every user and in
every viewing conditions (\emph{e.g.}, viewing distance, ambient light).
It may then be interesting to back up the VAC with another type of
measure than sporadic and disruptive questionnaires: EEG. EEG can be
measured in real-time without interrupting the viewing. EEG studies from
the literature do not take into account the VAC. Overall they compare
the state of the subjects before and after a long session of
stereoscopic watching \citep{Li2008a, Chen2013}; or they compare ``2D''
versus stereoscopy, without further control of what is displayed in
stereoscopy \citep{Li2008a, Cho2012, Chen2013}.

Therefore we decided to combine those two kinds of investigations to see
if the VAC could be supported with EEG recordings. Moreover, thanks to
precise stereoscopic parameters and real-time measures, one purpose of
this work is to build a protocol which compares short sequences of
several seconds instead of viewing sessions of several minutes (between
2min and 40min in \citep{Shibata2011, Li2008a, Chen2013}). If we are to
succeed, detecting comfort and no-comfort situations would allow in the
future to tune quickly stereoscopic display parameters for one
particular viewer.

\section{Materials and methods}\label{materials-and-methods}

We decided to study two different kinds of EEG signals: event-related
potentials (ERP) and frequency bands power. Since we wanted to avoid
bias and focus uniquely on the VAC, we chose to use a simple 3D scene: a
still gray cube over a black background. The only manipulated variable
was its apparent depth. With identical visual stimuli (shape, size,
colors) projected at each trial onto the retina, we restrained as much
as possible the signals that could modify EEG besides the VAC.\footnote{\begin{figure}[htbp]
  \centering
  \includegraphics{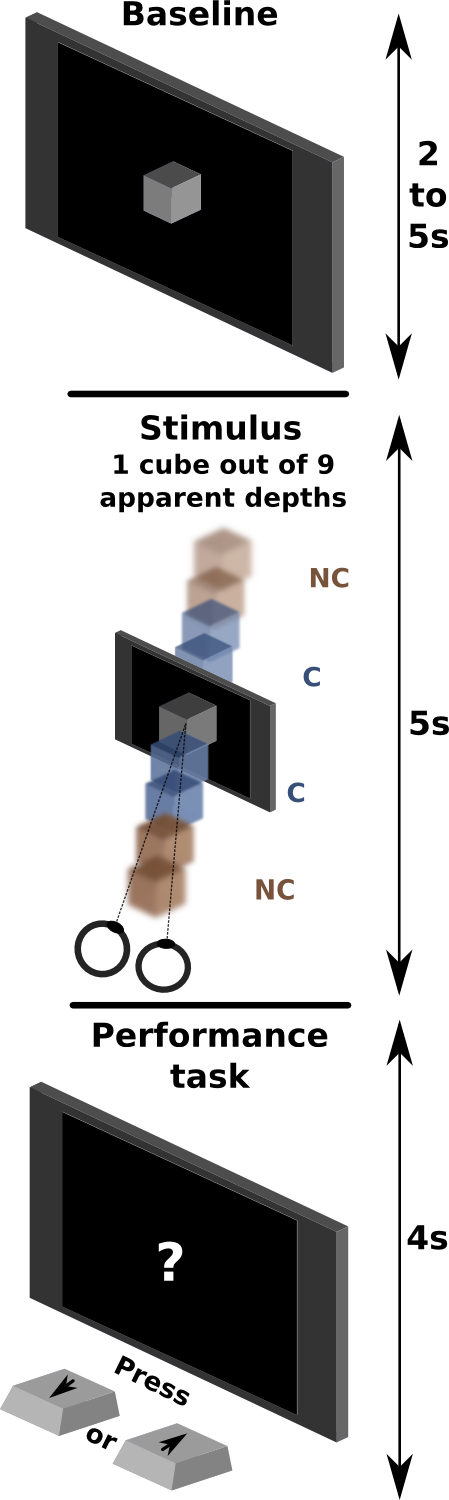}
  \caption{The steps of a trial: 2D baseline, gray cube at random depth,
  performance task.\label{trial_fig}}
  \end{figure}}

\subsection{Protocol}\label{protocol}

The experiment took place in a dedicated room with dimmed lights and a
quiet environment. Subjects were seated 1m away from an active display
of 65 inches (Panasonic TX-P65VT20E). We presented the cube at 9
different apparent depths during the experiment. 4 images appeared to be
in front of the screen, 4 behind and one was flat. We created our 8
stereoscopic images so as 4 laid inside the zone of comfort and 4
outside. Thus we defined a within-subject design with two conditions:
comfort (C) and no-comfort (NC). For the subjects the 9 cubes appeared
to be away at: 0.212m (NC); 0.424m (NC); 0.636m (C); 0.848m (C); 1m
(2D/``flat''); 1.512m (C); 2.023m (C); 2.534m (NC); 3.046m (NC).

Both for the purpose of maintaining attentive subjects during a tedious
experiment and for evaluating their ability to discriminate apparent
depths, we added a performance task. After each cube a question mark
appeared on the screen. Subjects were instructed to press at this moment
the ``up'' key if they believed the cube was ``behind the screen'', the
``down'' key if it was ``in front of the screen'' and nothing if they
couldn't decide. Their fingers were positioned prior to the session to
minimize motor efforts and prevent them to loose sight of the screen.

One trial consisted in 3 steps. First the 2D cube appeared on screen. It
served as a reference image. It stayed on screen during 2 to 5s so
subjects could not anticipate the appearance of the stereoscopic cues.
Then it was randomly replaced by 1 of the 9 possible cubes. This new
image remained 5s before the question mark appeared. During the 4s that
followed, the first key press -- if any -- was recorded. Then a new
trial immediately started. See figure \ref{trial_fig}. Each apparent
depth was presented 30 times. The overall 270 trials were balanced
across 5 sub-sessions.

The entire experiment comprised several sequences. Subjects entered in
the room. They were given general facts about the context of the
experiment and filled an informed consent form. To check their aptitude
to perceive stereoscopic images they completed a TNO test
\citep{Momeni-Moghadam2012}. If the test was successful, we proceeded to
the setup of the EEG device. Then subjects responded to the pre-test
questionnaire. We presented randomly the 9 images. After each image we
asked them to rate the clarity of their vision and the tiredness of
their eyes on a 5-point Likert scale; ``1'' representing no negative
symptoms and ``5'' severe symptoms. Those questions were adapted from
\citep{Shibata2011} and translated to French. They let us measure
respectively how well subjects saw the stereoscopic images and how
comfortable they felt. Before we started the main part of the
experiment, we accustomed them to the task with a training session. Once
they felt confident we began the first sub-session. On average a
sub-session lasted 12min. We kept on with the 4 others, letting subjects
rest a few minutes and drink water if they wanted to in-between. After
the sub-sessions were over we repeated post-test the Likert scale
questionnaires. Overall subjects stayed for about 2.5 hours in the room.

\subsection{EEG processing}\label{eeg-processing}

Three subjects took part in this pilot study (2 females, 1 male, mean
age 22, normal vision). Their EEG activity was recorded at a 512Hz
sampling rate with a g.tec g.USBamp system. 32 active electrodes were
positioned according to the international 10-20 system (reference on the
left earlobe). 4 were recording electrooculographic (EOG)
activity\oldfootnote{LO1, LO2, IO1 and FP1 sites} and 28 EEG
activity\oldfootnote{AF3, AF4, F7, F3, Fz, F4, F8, FC5, FC1, FC2, FC6, C3, Cz, C4, CP5, CP1, CP2, CP6, P7, P3, Pz, P4, P8, PO3, PO4, O1, Oz and O2 sites}.\footnote{\begin{figure}[htbp]
  \centering
  \includegraphics{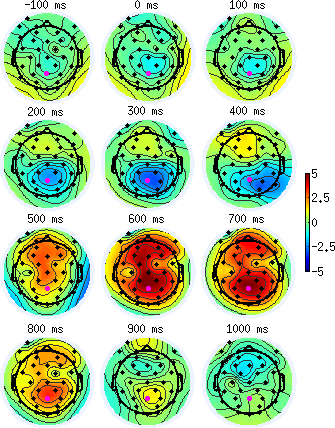}
  \caption{Scalp maps of ERP amplitudes over time (3 subjects, $\approx$
  240 trials over C and NC). Pz (highlighted in pink) is one of the site
  with the largest amplitude during stereoscopic viewing
  phase.\label{all_ERP_fig}}
  \end{figure}} OpenViBE \citep{OpenVibe} was used both to record
signals and to manage the experiment. EEGLAB \citep{Eeglab} 13.0.1b has
been used in conjunction with Matlab R2012B for EEG processing. First we
concatenated datasets from the 5 sub-sessions. We cleaned-up our signals
with a 0.5-25Hz band-pass filter. Then we extracted $9 \times 30 = 270$
epochs related to the cube appearance, -1s to +4s around stimulus onset.
Because of the unusual nature of the recordings -- few EEG studies
involve stereoscopy -- we chose automated artifacts rejection methods.
Epochs polluted with muscular activity were rejected with the EEGLAB
function pop\_autorej. Supported by \citep{Ghaderi2013}, we removed EOG
activity with the ADJUST toolbox. We ran an Infomax independent
component analysis on each dataset. We rejected components detected
either as eye blinks, vertical or horizontal eye movements by ADJUST.
Finally we compared subjects by pairing C and NC epochs.\footnote{\begin{figure}[htbp]
  \centering
  \includegraphics{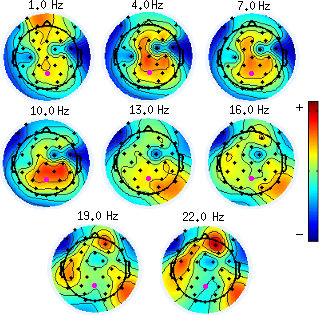}
  \caption{Scalp map of various frequencies' signal power (3 subjects,
  $\approx$ 240 trials over C and NC). Pz (highlighted in pink) is one
  of the site with the largest power during stereoscopic viewing
  phase.\label{all_freq_fig}}
  \end{figure}}

\subsection{Hypotheses}\label{hypotheses}

Consequently to this protocol design, we made 4 hypotheses.
\textbf{H1}:~The symptoms reported in Likert-scale questionnaires would
be more pronounced in the no-comfort condition than in the comfort one.
\textbf{H2}:~performance in the task would decrease in the NC condition
because of images more difficult to perceive. \textbf{H3}:~Similarly to
\citep{Li2008a}, delay in ERP would increase in the NC condition.
\textbf{H4}:~Frequency bands power would change between C and NC. In
particular the alpha activity would increase in NC because of more
disturbing stimuli.

\section{Results}\label{results}

We analyzed ERP from -0.5s to +2s around stimuli onsets and frequency
bands from -0.5s to +3.5s, averaging data from all subjects. Although in
the future more electrodes will have to be considered, in a first
approach this pilot study focuses on Pz. It is a standard site studied
in the literature and it is less likely to be polluted by EOG activity
than frontal sites. Moreover, for both ERP and frequency bands power,
scalp maps showed that Pz is among the most active sites in our data
(see figures \ref{all_ERP_fig} and \ref{all_freq_fig}). We chose to use
event-related spectral perturbations (ERSP) to gather first insights of
modifications in frequency bands power between C and NC. Statistical
analyses not related to EEG were performed with R 2.15.2. Because of the
small size of our samples we used non-parametric tests to analyze our
data -- permutation statistics in EEGLAB.

\textbf{H1}: A Wilcoxon Signed-rank test showed a significant effect of
the C/NC condition for both Likert-scale questionnaires (p \textless{}
0.01). Subjects reported more eye comfort in C than in NC (means: 1.83
\emph{vs} 2.67) and more vision clarity in C than in NC (means: 1.58
\emph{vs} 2.58).

\textbf{H2}: There was 3 possible answers during the performance task.
``Good'' were mapped to ``1'', ``bad'' to ``-1'' and ``none'' to ``0''.
A Wilcoxon Signed-rank test showed that there is a significant effect of
the C/NC condition in the performance task (p \textless{} 0.01, 120
trials for each subject and condition). The overall mean scores are 0.44
in C and 0.74 in NC. We ran a post-hoc analysis on answer types. Because
we couldn't match one C trial with one NC trial (they were randomly
distributed), we chose a Chi-square test. It showed a significant effect
between C and NC both for the ``good'' and ``none'' answers -- p
\textless{} 0.01, adjusted for multiple comparisons with false discovery
rate (FDR). C/``good'': 48\%, NC/``good'': 79\%. C/``none'': 48\%,
NC/``none'': 16\%.\footnote{\begin{figure}[htbp]
  \centering
  \includegraphics{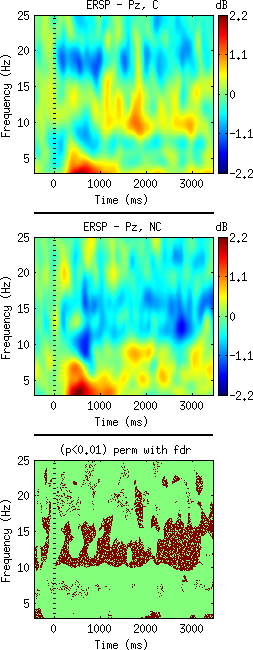}
  \caption{ERSP in Pz. \textbf{Top}: comfort condition; \textbf{middle}:
  no-comfort condition ($\approx$~120 trials each, 3 subjects).
  \textbf{Bottom}: red dots indicate significant differences in
  frequency powers between C and NC (p \textless{} 0.01, FDR
  correction).\label{Pz_ERSP_fig}}
  \end{figure}}

\begin{figure}[htbp]
\centering
\includegraphics{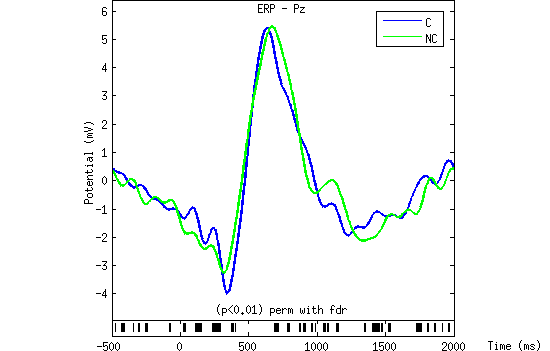}
\caption{Average ERP in Pz of comfort (blue) and no-comfort (green)
conditions ($\approx$~120 trials each, 3 subjects). Bottom marks
indicate significant differences (p \textless{} 0.01, FDR correction).
For easier interpretations a 8Hz low-pass filter is applied (view
only).\label{Pz_ERP_fig}}
\end{figure}

\textbf{H3}: Average ERP at Pz in C and NC showed significant
differences (p \textless{} 0.01, corrected with FDR). Figure
\ref{Pz_ERP_fig} suggests a delayed peak for the positive component in
NC. The differences appear more pronounced in the negative component
which precedes: greater amplitude in C.

\textbf{H4}: ERSP in Pz between C and NC showed significant differences
in several bands power (see figure \ref{Pz_ERSP_fig}). The most notable
is a \emph{decrease} within alpha band (10-14Hz) in NC; but an
\emph{increase} within theta (4-7Hz) and beta (15-25Hz) also seems to
occur in NC compared to C.

\section{Conclusion}\label{conclusion}

The main purpose of our study was to compare the EEG activity between
comfortable (C) and uncomfortable (NC) vergence-accommodation conflicts
in stereoscopic displays. Beside a better understanding of the
repercussions of this conflict within the brain, one goal would be
ultimately to build an adaptive system. Lightweight and practical EEG
devices could help to calibrate the display to each viewer, avoiding
uncomfortable experiences. In the literature subjects are mostly exposed
to long sessions of stereoscopic viewing before measures are performed.
Because an efficient adaptive system could not wait half an hour to find
the right parameters, we built a pilot study which relied instead on
short sequences of several seconds. We used a simple 3D scene in order
to limit bias.

According to the questionnaires, we replicated two different zones of
comfort as described in \citep{Shibata2011} (H1). We conceived a
performance task to sense how well subjects perceived depths.
Surprisingly, they performed \emph{better} in NC (H2). They did not give
a greater number of wrong answers in C though, instead they couldn't
decide more often if objects appeared in front of or behind the screen.
As such, the reported poorer vision clarity in NC does not seem to
correlate with a poorer depth perception. We remain cautious on our
interpretation though, since the cube we used lacked many cues which
should change the scene comprehension (e.g., shadows, relative
movements).

With EEG some results were similar to \citep{Li2008a} and
\citep{Cho2012}, even though we did not use an oddball paradigm and
compared two stereoscopic conditions instead of stereoscopic and 2D
images. In NC the ERP differed in their negative component (lower
amplitude) and their positive component (higher latency) (H3).
Unexpectedly, the frequency power in the alpha band was greater in C
than in NC (H4). Maybe our short sequences did not induce a greater
fatigue in NC, but instead made our subjects more attentive to the
stimuli. This result is similar to \citep{Cho2012} and \citep{Chen2013}.
However the increase in the beta band contradicts both and supports
\citep{Li2008a} instead. We stress again that these studies compared
stereoscopy to 2D. We should use more complex protocols and analyses to
understand exactly how EEG bands power relates to the VAC.

We managed to build a pilot protocol which supports the study of the VAC
through EEG. We presented a new method comparing different apparent
depths. There is several ways to improve on our study. We did not have
the material to measure pupil distance to reduce inter-subject
variability. On the same note, using a device such as the Oculus Rift
could give a far better accuracy in stereoscopy. We used simple stimuli;
random objects and orientations would allow us to consider the ``2D''
condition and create more realistic scenarios. If we show at the same
time objects with various depths, as in every movie, is it less
comfortable for the viewer? Fatigue and comfort goes beyond the VAC. We
chose one viewing distance but there is more to explore in the zone
comfort of \citep{Shibata2011}. We had to make two clusters (C/NC) from
our 8 stereoscopic images to gather enough EEG data. If we find a way to
present more stimuli while keeping the duration of the experiment
acceptable for subjects, more conditions could be studied.

When we pursued our analyses, we sensed that the ability to perceive
stereoscopy well could impact brain activity. EEG may not only be used
as a way to make stereoscopy comfortable, but to measure if it
\emph{works}. Moreover, when we added a condition ``in front
of''/``behind'' the screen, various profiles appeared. Whenever it is
for physiological or cognitive reasons, we may not perceive equally both
effects. Once the protocol is strengthened, more subjects will have to
be involved so as to draw strong conclusions.

By extending our protocol it should be possible to study at the same
time comfort and depth perception. With more data, we could build an EEG
classifier and also have a better understanding of stereoscopy:
guidelines for an improved technology and greater entertainments.

\bibliography{biblio.bib}

\end{document}